\documentclass[sigconf, natbib=true]{acmart}

\AtBeginDocument{%
  }
\usepackage{hyperref}
\usepackage{multirow}
\usepackage{multicol}
\usepackage{caption}
\usepackage{subcaption}
\usepackage{latexsym}
\usepackage{mflogo}
\usepackage{graphics}
\usepackage{tabularx}
\setcopyright{acmcopyright}
\usepackage{threeparttable}
 
\newcommand{\ignore}[1]{}
\usepackage{xspace}

\newcommand{\ie}{\emph{i.e.,}\xspace}

\newcommand{\eg}{\emph{e.g.,}\xspace}

\newcommand{\ourname}{KuaiSAR\xspace}

\copyrightyear{2023} 
\acmYear{2023} 
\setcopyright{acmlicensed}\acmConference[CIKM '23]{Proceedings of the 32nd ACM International Conference on Information and Knowledge Management}{October 21--25, 2023}{Birmingham, United Kingdom} \acmBooktitle{Proceedings of the 32nd ACM International Conference on Information and Knowledge Management (CIKM '23), October 21--25, 2023, Birmingham, United Kingdom} 
\acmPrice{15.00} 

\begin{document}

\title{KuaiSAR: A Unified Search And Recommendation Dataset}


\author{Zhongxiang Sun}
\authornote{Equal Contribution.
Work done during their internships at Kuaishou.}
\author{Zihua Si}
\authornotemark[1]
\affiliation{
  \institution{Gaoling School of Artificial
Intelligence\\Renmin University of China}
  \city{Beijing}\country{China}
  }
\email{{sunzhongxiang, zihua_si}@ruc.edu.cn}

\author{Xiaoxue Zang}
\author{Dewei Leng}
\affiliation{%
  \institution{Kuaishou Technology Co., Ltd.}
  \city{Beijing}\country{China}
  }
\email{{zangxiaoxue, lengdewei}@kuaishou.com}

\author{Yanan Niu}
\author{Yang Song}
\affiliation{%
  \institution{Kuaishou Technology Co., Ltd.}
  \city{Beijing}\country{China}
  }
\email{{niuyanan, yangsong}@kuaishou.com}

\author{Xiao Zhang}
\author{Jun Xu}
\authornote{Jun Xu is the corresponding author. Work partially done at Engineering Research Center of Next-Generation Intelligent Search and Recommendation, Ministry of Education.}
\affiliation{%
  \institution{Gaoling School of Artificial
Intelligence\\Renmin University of China}
  \city{Beijing}\country{China}
  }
\email{{zhangx89, junxu}@ruc.edu.cn}

\renewcommand{\shortauthors}{Zhongxiang Sun et al.}

\begin{abstract}
The confluence of Search and Recommendation (S\&R) services is vital to online services, including e-commerce and video platforms.
The integration of S\&R modeling is a highly intuitive approach adopted by industry practitioners.
However, there is a noticeable lack of research conducted in this area within academia, primarily due to the absence of publicly available datasets. 
Consequently, a substantial gap has emerged between academia and industry regarding research endeavors in joint optimization using user behavior data from both S\&R services.
To bridge this gap, we introduce the first large-scale, real-world dataset \textbf{\ourname} of integrated \emph{S}earch \emph{A}nd \emph{R}ecommendation behaviors collected from \emph{Kuai}shou, a leading short-video app in China with over 350 million daily active users.  
Previous research in this field has predominantly employed publicly available semi-synthetic datasets~\cite{aiqingyao_search,IV4REC}, with artificially fabricated search behaviors.
Distinct from previous datasets, \ourname contains genuine user behaviors, including the occurrence of each interaction within either search or recommendation service, and the users' transitions between the two services. 
This work aids in joint modeling of S\&R, and utilizing search data for recommender systems (and recommendation data for search engines). 
Furthermore, due to the various feedback labels associated with user-video interactions, \ourname also supports a broad range of tasks, including intent recommendation, multi-task learning, and modeling of long sequential multi-behavioral patterns. 
We believe this dataset will serve as a catalyst for innovative research and bridge the gap between academia and industry in understanding the S\&R services in practical, real-world applications.
The dataset is available at \textcolor{magenta}{\url{https://ethan00si.github.io/KuaiSAR/}}.
The dataset is also shared at \textcolor{magenta}{\url{https://zenodo.org/record/8181109}}.

\end{abstract}

\ccsdesc[500]{Information systems~Information retrieval}


\keywords{Datasets; Recommendation; Search}

\maketitle

\section{Introduction}

Many online content platforms, \eg Kuaishou and TikTok, provide search and recommendation (\textbf{S\&R}) services to satisfy user's diverse information needs.
The differentiation between S\&R services may be absent to users on these platforms.
The integration of S\&R services is a highly intuitive approach adopted by industry practitioners.
However, academic research is scarce in this domain, primarily due to the lack of publicly available datasets that support scholarly investigations. 
Consequently, a substantial gap has emerged between academia and industry regarding research endeavors in this particular domain.

The lack of large-scale, real-world datasets on user S\&R behaviors has limited the research progress of joint modeling of S\&R. 
Existing datasets in the field of recommendation systems (or personalized search) only contain user behavior sequences within the recommendation system (or search engine).
Previous research in this field has usually depended on experiments conducted using proprietary industrial datasets~\cite{USER, SRJgraph, Intent_Reco_ali_CIKM21} or simulated datasets~\cite{JSR2,IV4REC,JSR}, thereby hindering the participation of more researchers.

\begin{figure}
    \centering
        \includegraphics[width=0.99 \linewidth]{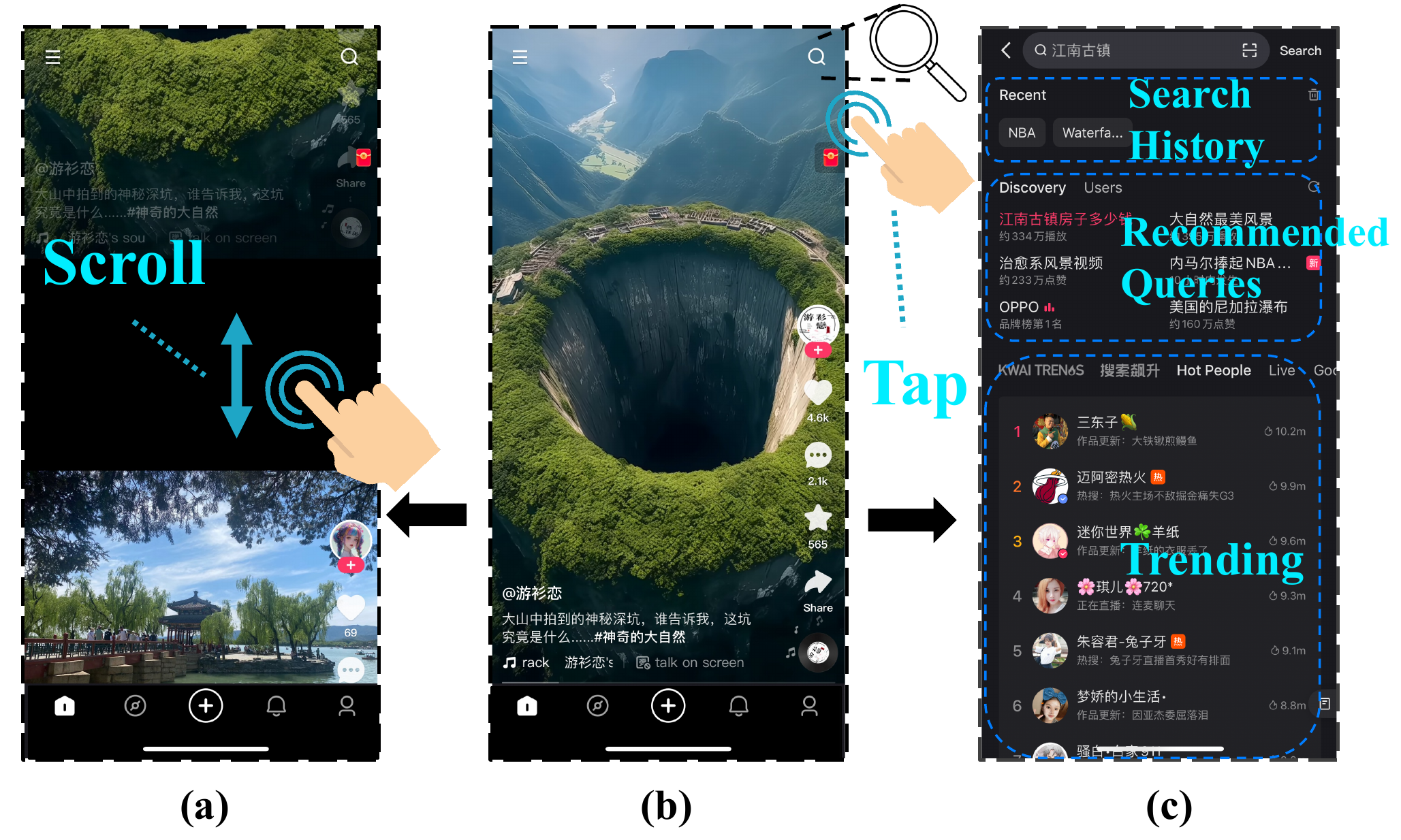}
    \vspace{-5pt}
    \caption{The integrated S\&R scenarios in Kuaishou app. 
    In the process of watching a video, users have two primary interaction modalities. They can either utilize the recommendation service, involving scrolling vertically to discover diverse videos (b) -> (a). Alternatively, they can tap the magnifying glass symbol to leverage the search service (b) -> (c).
    }
\label{fig:model}
\end{figure}

\textbf{Existing public datasets suffer from the following two deficiencies:}
(1) There is no dataset that collects the behavioral history of users in both the recommender system and search engine.
(2) The previous datasets are not based on a unified scenario that offers interactive S\&R services.
In this scenario, the search service integrates elements of recommendation, while the recommendation service is closely intertwined with the search functionality.
Numerous mobile applications have developed such integrated scenarios in recent years, exemplified by Kuaishou\footnote{https://www.kuaishou.com/en}. As depicted in~\autoref{fig:model}, users may transition to the search interface while utilizing the recommendation system, and they may also encounter recommended queries while using the search engine (details in~Section~\ref{sec:Kuaishou details}).

\textbf{The contributions of \ourname are summarized as follows:}
To facilitate academic research in exploring the potential of integrating S\&R services, we present a large-scale real-world dataset containing both S\&R behaviors, called \ourname. \ourname is collected from the Kuaishou app, one of China's largest short-video apps with more than 350 million daily active users.  
\autoref{tab: statistics} presents basic statistics of \ourname.
In contrast to the previous datasets, \ourname contains rich information: First, it contains users \emph{genuine} S\&R behaviors, and explicitly records the \emph{user-system interactions} in both the recommendation service and the search service.
Moreover, it records whether users \emph{transition} to the search service through the current video while using the recommendation service. 
Finally, it captures the sources of user entry into the search service, \eg actively typing a query and clicking on a recommended query.

\begin{table}[t]
 \caption {Statistics of user actions in \ourname (top) and feature descriptions (bottom). `S’ and `R’
denote search and recommendation, respectively.
All users have both S\&R behaviors. 
}\label{tab: statistics} 

\tabcolsep=7.5pt 
\begin{tabular}{l|cccc}
\toprule
 Dataset & \#Users & \#Items & \#Queries &  \#Actions    \\ \hline
 S-data & 25,877  & 3,026,189 & 453,667   & 5,059,169   \\
 R-data & 25,877  & 4,046,367 & - & 14,605,716 \\ 
 Total & 25,877  & 6,890,707 & 453,667  & 19,664,885 \\ 
 \bottomrule
\end{tabular}

\vspace{2pt}

\tabcolsep=7.5pt
\begin{tabular}{@{}ll@{}}
\toprule
\textbf{User\&Item feature}:  & \begin{tabular}[c]{@{}l@{}} Users and items have abundant side\\ information. $5$ ($18$) features for users\\ (items). \end{tabular}                                                                                   \\ \midrule
\textbf{S-action feature}:  & \begin{tabular}[c]{@{}l@{}} S-actions have $9$ features, \eg  search\\ session IDs, query keywords, and\\ sources of entering the search service. \end{tabular}                                                                                   \\ \midrule
\textbf{R-action feature}:  & \begin{tabular}[c]{@{}l@{}} R-actions has $12$ features, including $9$\\ types of user feedback, \eg likes,\\ follows, and entering search.\end{tabular}                                                                                   \\ \midrule
\textbf{Social network}: & \begin{tabular}[c]{@{}l@{}}  576 users have friends.\end{tabular} \\ \bottomrule
\end{tabular}
\end{table}

It is noteworthy that \ourname is the first dataset that records genuine user S\&R behaviors within an interactive app that provides unified search and recommendation services.
This dataset has the potential to advance the research of joint modeling for S\&R~\cite{JSR, JSR2, SRJgraph, USER}, facilitate better utilization of search data in recommendation systems~\cite{IV4REC,IV4Rec+,SESRec,NRHUB}, utilization of recommendation data in search~\cite{aiqingyao_search, personalized_search_SIGIR20}, as well as the intent recommendation~\cite{Intent_Reco_ali_CIKM21, intent_reco_KDD19}.
Given the diverse labels for user-video interactions in the dataset, \eg whether search, like, and forward, \ourname can also facilitate a wide range of tasks, including multi-task learning~\cite{MMOE, PLE} and multi-behavior sequential modeling~\cite{Multi_behavior_SIGIR22,multi_behavior_SIGIR20}.

\section{Related work}
\subsection{Joint Search and Recommendation}
An early study~\cite{garcia2011information} pointed out that search (information retrieval) and recommendation (information filtering) are the two sides of the same coin.
These two services share similar objectives~\cite{xu2018deep}, which are to provide users with information to fulfill their needs. The key distinction lies in whether the user's needs involve explicit queries.

Recently, many studies have recognized the potential for joint modeling of S\&R.
Several studies~\cite{SRJgraph,USER} propose the design of a unified model that integrates S\&R, effectively modeling user interests.
Some works~\cite{JSR,JSR2} have devised joint loss functions to train S\&R models simultaneously.
Another research direction involves utilizing the behavioral data from one service to assist in modeling the other service. 
This direction entails leveraging search data to enhance recommendation models~\cite{NRHUB,IV4REC,SESRec} or employing recommendation data to augment search models~\cite{Liu_www22,aiqingyao_search}.
Some industry practitioners have also realized the potential value of integrating S\&R.
Consequently, in practical scenarios, many search behaviors are initiated through the recommendation system. This particular type of recommendation service focused on recommending search queries, is commonly referred to as ``Intent Recommendation''~\cite{intent_reco_KDD19, Intent_Reco_ali_CIKM21, muliti_scenario_SIGIR22}.

However, existing research in this field has primarily relied on private datasets~\cite{SRJgraph,USER, intent_reco_KDD19,NRHUB} or semi-synthetic public datasets~\cite{JSR,JSR2}. The lack of a large-scale, real-world dataset encompassing both S\&R behaviors has constrained the advancement of this field.

\subsection{Existing Dataset}
The currently available search or recommendation datasets have predominantly been designed to cater to a single research field, focusing either on search or recommendation.
For instance, MS MARCO~\cite{Xiong2019OpenDW} provides query and document information along with query-document interactions for research in the search domain. 
KuaiRec~\cite{gao2022kuairec}, on the other hand, offers user-item interactions in the recommendation service, catering to research in the recommendation field.
Only a few datasets attempt to provide user S\&R behaviors concurrently.
To the best of our knowledge, only two existing datasets provide both search and recommendation interactions.

\begin{table}[!t]
    \tabcolsep=1.5pt
    \renewcommand\arraystretch{0.9}
    \caption{Comparison of currently available datasets for search and recommendation.}
    \label{tab:datasets}
    \begin{threeparttable}[b]
    \begin{tabular}{l|cccccc}
    \toprule
    \textbf{Property} & \textbf{MS MARCO} & \textbf{KuaiRec} & \textbf{Amazon} & \textbf{JDsearch} & \textbf{KuaiSAR} \\ \hline
    \textbf{R-action}                                                     & No                          & Yes                          & Yes                          & Yes\tnote{2}                          & Yes                          \\ 
    \textbf{S-action}                                                     & Yes                   & No                   & Yes\tnote{1}       & Yes                  & Yes                  \\ 
    \textbf{\# Users}                                                   & --                               & 7,176                               & 192,403                               & 173,831                               & 25,877                                \\ 
    \textbf{\# Items}                                                   & 8,841,823                               & 10,728                                & 63,001                                & 12,872,636                            & 6,890,707                             \\ 
    \textbf{\# Queries}                                                & 509,919                                 & --                                 & 3,221                                  & 171,728                                   & 453,667                                \\ 
    \textbf{\# Actions}                                                & 509,919                           & 17,207,376                           & 1,689,188                            & 26,667,260                           & 19,664,885                          \\ 
    \bottomrule
\end{tabular}
\begin{tablenotes}
    \item[1] Search actions in the Amazon dataset are artificially simulated.
    \item[2] The JDsearch dataset lacks distinction between non-search data originating from recommendation scenarios or casual browsing.
\end{tablenotes}
\end{threeparttable}
\end{table}

We briefly introduce these two existing datasets.
One is a widely used semi-synthetic dataset, while the other is a recently released real-world dataset.

$\bullet$ \textbf{Amazon}~\cite{amazon_dataset, aiqingyao_search}. This dataset was initially created for recommendation systems~\cite{amazon_dataset}. It consists of user review data and item metadata extracted from purchases made on Amazon.
Some researchers~\cite{aiqingyao_search} have utilized item metadata to construct pseudo queries, simulating user search behaviors and thereby enabling the dataset to encompass both user S\&R behaviors.
Due to the lack of publicly available data, this dataset has ultimately become the most commonly used dataset in the field of joint modeling of S\&R~\cite{JSR,JSR2,Liu_www22,SESRec,personalized_search_SIGIR20}.
The obvious drawback of this dataset is the lack of real user search behaviors.

$\bullet$ \textbf{JDsearch}~\cite{liu2023JDsearch}. 
It is a recently released dataset designed specifically for personalized product search.
It consists of user queries and diverse user-product interactions collected from JD.com, a popular Chinese e-commerce platform.
The interactions may be from diverse channels, as stated in~\cite{liu2023JDsearch}, including ``search, recommendation, and casual browsing''.
The limitation of this dataset lies in its mere differentiation between non-search data, without capturing whether non-search data is in recommendation scenarios or the casual browsing scenario. Additionally, it does not document whether search interactions come from typing-in searches or clicking on recommended queries.

We compared \ourname with the datasets above, and the comparison results are listed in~\autoref{tab:datasets}.
Compared with the existing datasets, \ourname has the following key advantages: 
(1) it records and discriminates between users' authentic search and recommendation behaviors;
(2) it documents the sources of users' search behaviors, \eg actively typing in searches and clicking on recommended queries;
(3) it comprehensively captures users' transitions between S\&R services, such as documenting whether users initiate a search while watching a video within the recommendation system;
(4) it provides abundant side information for both users and items; and
(5) it logs users' authentic interactions, including both positive and negative feedback.

\section{Dataset Description}

\subsection{Characteristics of Kuaishou App}
\label{sec:Kuaishou details}


Kuaishou is one of the most popular short video-sharing platforms in China, with over 350 million daily active users.
As shown in~\autoref{fig:model}, Kuaishou provides both S\&R services.
As users scroll down the screen, they can discover new recommended short videos they may be interested in.
When users click on the magnifying glass, they can enter the main search page and use the search engine to find videos of interest.

In recent years, Kuaishou has focused on integrating S\&R services to enhance user experiences. The recommender leverages query-based recommendations, such as suggesting queries in the comments section, to encourage users to explore new information and resources using the search service. 
Similarly, incorporating recommended queries on the main search page helps users discover content relevant to their interests, stimulating curiosity and prompting the exploration of more engaging material. For detailed examples, please visit our \href{https://ethan00si.github.io/KuaiSAR/}{website}. 
In summary, the collaborative efforts of these systems improve information delivery and enhance the user experience.

Considering the specific characteristics of the Kuaishou, we have introduced additional labels in the user behavior logs to foster potential research.
These labels aim to  capture the transitions occurring between S\&R services accurately. 
For user recommendation behaviors, we record whether users tap on the magnifying glass to search while browsing videos within the recommendation system, as well as whether these queries are related to the current video.
For user search behaviors, we record the sources of their entry into the search engine, such as clicking on recommended related queries, manually typing queries, and clicking on hot search topics.
These labels can enhance our understanding of user behaviors  within S\&R services on a more comprehensive level.

\subsection{Data Construction}
To facilitate the research on integrating S\&R services, \ourname is constructed with the following steps:

     First, we randomly sampled approximately 25,000 users who accessed both search and recommendation services in Kuaishou app between May 22, 2023 and June 10, 2023. 
     The user interaction behaviors 
     include \emph{not only the positive feedback but also the negative feedback}. 
     For instance, in the recommendation scenario, negative interactions include videos that were displayed but skipped by the users; In the search scenario, the negative interactions involve the search results that were exposed but not clicked by the users.
     In addition, \emph{various user feedback} in the recommendation system is also recorded, including likes, shares, follows, and playing time. 
     Considering that users may switch between search and recommendation services, we also capture whether users initiate a search while viewing a video (labeled as `search') and whether the search query is related to the current video (labeled as `search\_photo\_related'), as two additional labels.
     The diverse user feedback labels provide an opportunity to investigate the interest transition of users in the unified S\&R scenario.
     The timestamps of these actions were also recorded in the dataset, providing temporal information for models. 
     Moreover, the data is clearly documented, specifying their respective occurrence scenarios, \ie whether in the search or in the recommendation scenario.
     We also included users' social network information, which can be used to enrich research on social networks using comprehensive S\&R data.
     Considering that users may actively initiate a search or enter the search service by clicking on triggered terms in recommendation, the sources of user search interactions are meticulously recorded, allowing for differentiation of different types of search behaviors.

    Second, we collected various side information for users and items.
    As for items, informative features include captions, author ID, photo types, uploading date, uploading type, music ID, topic tags, and category types of four levels.
    As for users, we recorded their activity levels in both search and recommendation services. We also included two encrypted features for each user.

    Finally, we anonymized the collected records to protect privacy according to the data-releasing policy. The ids of videos, users, and other entities were randomly hashed into integers.
    Textual information, such as queries and video titles, underwent word segmentation and sensitive word removal. Furthermore, terms in texts were randomly hashed into integers. 
    The anonymization safeguards the dataset against any inclusion of personal and private information while maintaining its integrity.

\subsection{Statistics}

\ourname contains genuine S\&R behaviors of 25,877 users within a span of 19 days on the Kuaishou app.
The basic statistics of \ourname are summarized in~\autoref{tab: statistics}.
For more specific statistical data and usage, please refer to \url{https://ethan00si.github.io/KuaiSAR/}.

\begin{figure}
    \centering
    \begin{subfigure}{0.49\linewidth}
        \centering
        \includegraphics[width=\textwidth]{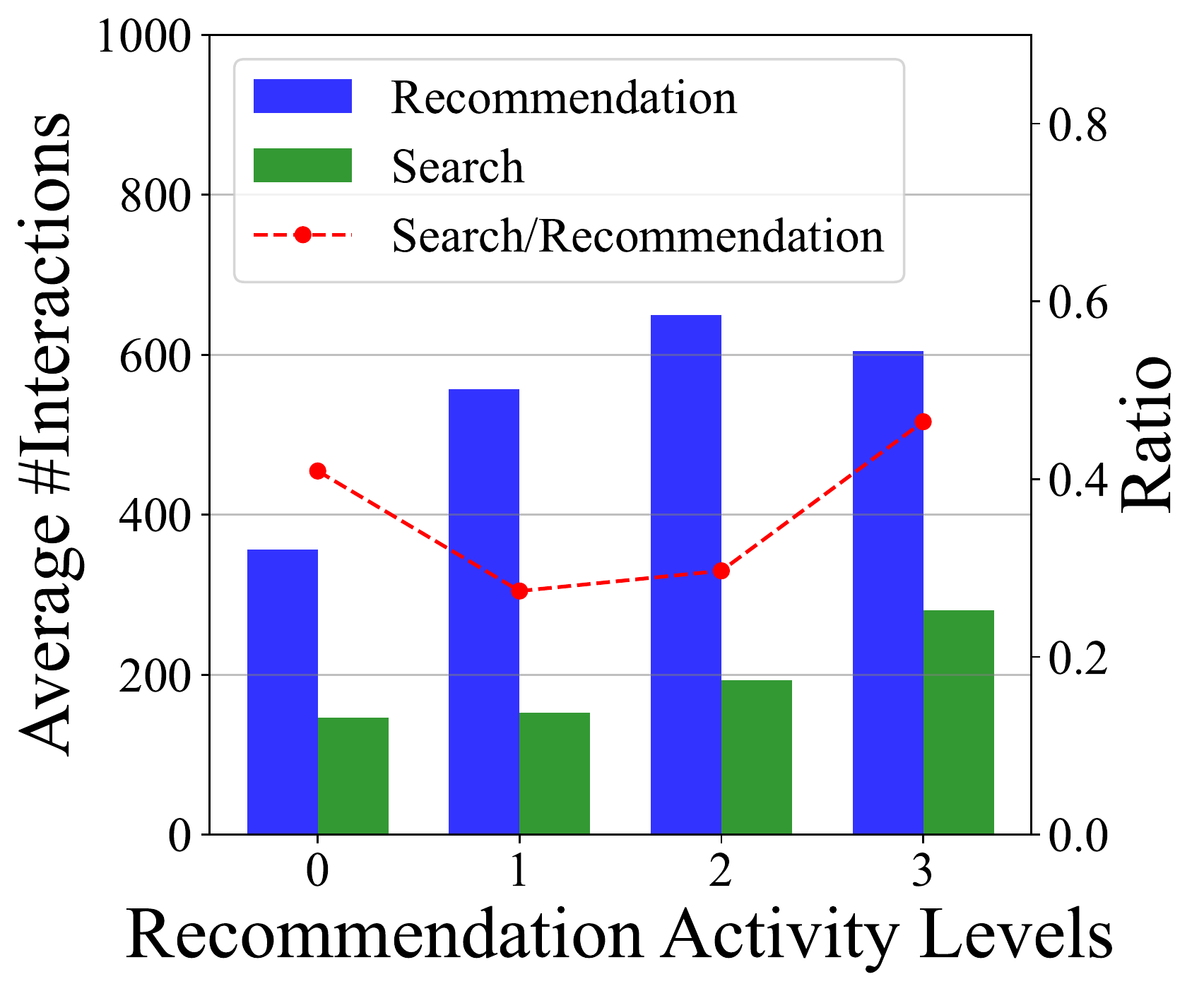}
        \subcaption{Users grouped by recommendation activity levels.}
    \end{subfigure}
    \begin{subfigure}{0.49\linewidth}
        \centering
        \includegraphics[width=\textwidth]{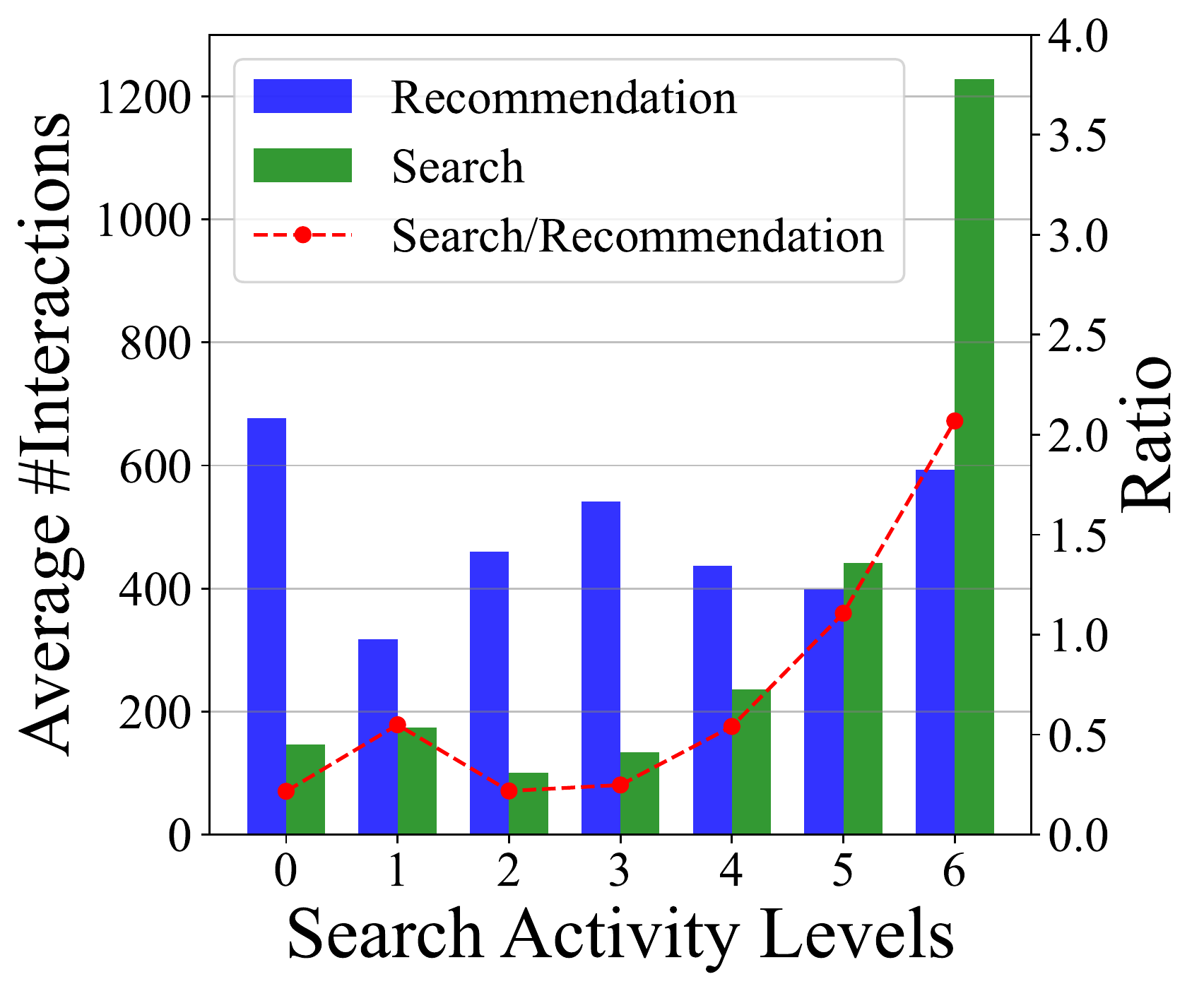}
        \subcaption{Users grouped by search activity levels.}
    \end{subfigure}
    \vspace{-2pt}
    \caption{Distribution of user interactions in S\&R scenarios.
    A higher activity level indicates a higher level of user engagement.
    The red line represents the ratio of the number of search interactions to the number of recommendation interactions.
    }
\label{fig:statistics}
\vspace{-10pt}
\end{figure}

This dataset filters users based on a single condition: users have used both S\&R services  within the specified time period.
As a result, the final dataset encompasses users with diverse levels of activity in either the search or recommendation services, thereby offering a comprehensive representation of users with varying degrees of engagement.
To illustrate the number of S\&R behaviors among users with different activity levels, we counted the number of user-video interactions within two services respectively.
We have grouped users based on their activity levels in the search or recommendation services. 
The activity level is determined by the number of active days within the past month using the respective service. 
A higher activity level indicates a larger number of active days.
The results are illustrated in~\autoref{fig:statistics}.
The average number of search or recommendation historical behaviors per user is over one hundred.  The overall interaction frequency with the recommendation service surpasses that with the search service. Furthermore, we observed that within the groups with either the lowest or highest activity levels in recommendations, as well as within the group with high search activity, there is a higher proportion of search interactions.

\section{Potential Research  Direction}
By open-sourcing \ourname, we provide an opportunity to propel the development and innovation of joint modeling for S\&R:

\noindent$\bullet$ \textbf{Unified Search and Recommendation.}
Previous works have proposed the approach of joint training to simultaneously optimize S\&R models~\cite{JSR, JSR2}, or employing a unified model to provide both services concurrently~\cite{USER, SRJgraph}.
\ourname is the pioneering one that provides authentic user behaviors in both services.

\noindent$\bullet$ \textbf{Enhanced Recommendation by Search (or Enhanced Search by Recommendation).}
It is reasonable and natural to employ one service to enhance the other.
The recommendation model can leverage search data to comprehensively understand user interests or item representations~\cite{IV4REC,SESRec,NRHUB}.
The search model can alleviate cold-start issues~\cite{Google_trasnfer_from_src_to_reco} or enable more precise personalized search~\cite{personalized_search_SIGIR20, Liu_www22} by incorporating recommendation data.

\noindent$\bullet$ \textbf{Intent Recommendation.}
In real-world applications, the recommendation model can also stimulate users to engage in more search behaviors by suggesting queries (so-called intent recommendation)~\cite{Intent_Reco_ali_CIKM21, intent_reco_KDD19, intent_reco_www21}.
\ourname captures how users initiate the search, such as by clicking recommended terms or clicking on related searches.

Given the abundant labels covering various types of user actions, \ourname can also unlock opportunities for several other promising research directions:

\noindent$\bullet$ \textbf{Multi-task Learning.}
S\&R, in essence, are different tasks designed.
\ourname provides an opportunity for multi-task learning~\cite{MMOE}, tailored for these two closely related yet distinct tasks. 
Furthermore, S\&R can also be seen as two scenarios, which can support research for multi-scenario models~\cite{muliti_scenario_SIGIR22}.


\noindent$\bullet$ \textbf{Sequential Multi-behavioral Modeling.}
In these years, there has been a growing interest in exploring how user modeling can be performed based on multiple types of user behaviors in sequential recommendation~\cite{Multi_behavior_SIGIR22,multi_behavior_www22, DMT_JD} or streaming recommendation~\cite{Zhang2021Counterfactual}.
\ourname  presents new research possibilities in sequential multi-behavioral modeling.

\begin{acks}
This work was funded by the National Key R\&D Program of China (2019YFE0198200), Kuaishou Technology, Beijing Outstanding Young Scientist Program NO. BJJWZYJH012019100020098, Intelligent Social Governance Interdisciplinary Platform, Major Innovation \& Planning Interdisciplinary Platform for the ``Double-First Class'' Initiative, Renmin University of China. Supported by fund for building world-class universities (disciplines) of Renmin University of China.
\end{acks}


\balance
\bibliographystyle{ACM-Reference-Format}
\bibliography{sample-base}


\begin{thebibliography}{29}


\ifx \showCODEN    \undefined \def \showCODEN     #1{\unskip}     \fi
\ifx \showDOI      \undefined \def \showDOI       #1{#1}\fi
\ifx \showISBNx    \undefined \def \showISBNx     #1{\unskip}     \fi
\ifx \showISBNxiii \undefined \def \showISBNxiii  #1{\unskip}     \fi
\ifx \showISSN     \undefined \def \showISSN      #1{\unskip}     \fi
\ifx \showLCCN     \undefined \def \showLCCN      #1{\unskip}     \fi
\ifx \shownote     \undefined \def \shownote      #1{#1}          \fi
\ifx \showarticletitle \undefined \def \showarticletitle #1{#1}   \fi
\ifx \showURL      \undefined \def \showURL       {\relax}        \fi
\providecommand\bibfield[2]{#2}
\providecommand\bibinfo[2]{#2}
\providecommand\natexlab[1]{#1}
\providecommand\showeprint[2][]{arXiv:#2}

\bibitem[Ai et~al\mbox{.}(2017)]%
        {aiqingyao_search}
\bibfield{author}{\bibinfo{person}{Qingyao Ai}, \bibinfo{person}{Yongfeng
  Zhang}, \bibinfo{person}{Keping Bi}, \bibinfo{person}{Xu Chen}, {and}
  \bibinfo{person}{W.~Bruce Croft}.} \bibinfo{year}{2017}\natexlab{}.
\newblock \showarticletitle{Learning a Hierarchical Embedding Model for
  Personalized Product Search}. In \bibinfo{booktitle}{\emph{Proceedings of the
  40th International ACM SIGIR Conference on Research and Development in
  Information Retrieval}} (Shinjuku, Tokyo, Japan)
  \emph{(\bibinfo{series}{SIGIR '17})}. \bibinfo{publisher}{Association for
  Computing Machinery}, \bibinfo{address}{New York, NY, USA},
  \bibinfo{pages}{645–654}.
\newblock
\showISBNx{9781450350228}
\urldef\tempurl%
\url{https://doi.org/10.1145/3077136.3080813}
\showDOI{\tempurl}


\bibitem[Bi et~al\mbox{.}(2020)]%
        {personalized_search_SIGIR20}
\bibfield{author}{\bibinfo{person}{Keping Bi}, \bibinfo{person}{Qingyao Ai},
  {and} \bibinfo{person}{W.~Bruce Croft}.} \bibinfo{year}{2020}\natexlab{}.
\newblock \showarticletitle{A Transformer-Based Embedding Model for
  Personalized Product Search}. In \bibinfo{booktitle}{\emph{Proceedings of the
  43rd International ACM SIGIR Conference on Research and Development in
  Information Retrieval}} (Virtual Event, China) \emph{(\bibinfo{series}{SIGIR
  '20})}. \bibinfo{publisher}{Association for Computing Machinery},
  \bibinfo{address}{New York, NY, USA}, \bibinfo{pages}{1521–1524}.
\newblock
\showISBNx{9781450380164}
\urldef\tempurl%
\url{https://doi.org/10.1145/3397271.3401192}
\showDOI{\tempurl}


\bibitem[Fan et~al\mbox{.}(2019)]%
        {intent_reco_KDD19}
\bibfield{author}{\bibinfo{person}{Shaohua Fan}, \bibinfo{person}{Junxiong
  Zhu}, \bibinfo{person}{Xiaotian Han}, \bibinfo{person}{Chuan Shi},
  \bibinfo{person}{Linmei Hu}, \bibinfo{person}{Biyu Ma}, {and}
  \bibinfo{person}{Yongliang Li}.} \bibinfo{year}{2019}\natexlab{}.
\newblock \showarticletitle{Metapath-guided Heterogeneous Graph Neural Network
  for Intent Recommendation}. In \bibinfo{booktitle}{\emph{Proceedings of the
  25th {ACM} {SIGKDD} International Conference on Knowledge Discovery {\&} Data
  Mining, {KDD} 2019, Anchorage, AK, USA, August 4-8, 2019}},
  \bibfield{editor}{\bibinfo{person}{Ankur Teredesai}, \bibinfo{person}{Vipin
  Kumar}, \bibinfo{person}{Ying Li}, \bibinfo{person}{R{\'{o}}mer Rosales},
  \bibinfo{person}{Evimaria Terzi}, {and} \bibinfo{person}{George Karypis}}
  (Eds.). \bibinfo{publisher}{{ACM}}, \bibinfo{pages}{2478--2486}.
\newblock
\urldef\tempurl%
\url{https://doi.org/10.1145/3292500.3330673}
\showDOI{\tempurl}


\bibitem[Gao et~al\mbox{.}(2022)]%
        {gao2022kuairec}
\bibfield{author}{\bibinfo{person}{Chongming Gao}, \bibinfo{person}{Shijun Li},
  \bibinfo{person}{Wenqiang Lei}, \bibinfo{person}{Jiawei Chen},
  \bibinfo{person}{Biao Li}, \bibinfo{person}{Peng Jiang},
  \bibinfo{person}{Xiangnan He}, \bibinfo{person}{Jiaxin Mao}, {and}
  \bibinfo{person}{Tat-Seng Chua}.} \bibinfo{year}{2022}\natexlab{}.
\newblock \showarticletitle{KuaiRec: A Fully-Observed Dataset and Insights for
  Evaluating Recommender Systems}. In \bibinfo{booktitle}{\emph{Proceedings of
  the 31st ACM International Conference on Information \& Knowledge
  Management}} (Atlanta, GA, USA) \emph{(\bibinfo{series}{CIKM '22})}.
  \bibinfo{pages}{540–550}.
\newblock
\urldef\tempurl%
\url{https://doi.org/10.1145/3511808.3557220}
\showDOI{\tempurl}


\bibitem[Garcia-Molina et~al\mbox{.}(2011)]%
        {garcia2011information}
\bibfield{author}{\bibinfo{person}{Hector Garcia-Molina},
  \bibinfo{person}{Georgia Koutrika}, {and} \bibinfo{person}{Aditya
  Parameswaran}.} \bibinfo{year}{2011}\natexlab{}.
\newblock \showarticletitle{Information seeking: convergence of search,
  recommendations, and advertising}.
\newblock \bibinfo{journal}{\emph{Commun. ACM}} \bibinfo{volume}{54},
  \bibinfo{number}{11} (\bibinfo{year}{2011}), \bibinfo{pages}{121--130}.
\newblock


\bibitem[Gu et~al\mbox{.}(2021)]%
        {Intent_Reco_ali_CIKM21}
\bibfield{author}{\bibinfo{person}{Yulong Gu}, \bibinfo{person}{Wentian Bao},
  \bibinfo{person}{Dan Ou}, \bibinfo{person}{Xiang Li},
  \bibinfo{person}{Baoliang Cui}, \bibinfo{person}{Biyu Ma},
  \bibinfo{person}{Haikuan Huang}, \bibinfo{person}{Qingwen Liu}, {and}
  \bibinfo{person}{Xiaoyi Zeng}.} \bibinfo{year}{2021}\natexlab{}.
\newblock \showarticletitle{Self-Supervised Learning on Users' Spontaneous
  Behaviors for Multi-Scenario Ranking in E-commerce}. In
  \bibinfo{booktitle}{\emph{{CIKM} '21: The 30th {ACM} International Conference
  on Information and Knowledge Management, Virtual Event, Queensland,
  Australia, November 1 - 5, 2021}},
  \bibfield{editor}{\bibinfo{person}{Gianluca Demartini},
  \bibinfo{person}{Guido Zuccon}, \bibinfo{person}{J.~Shane Culpepper},
  \bibinfo{person}{Zi~Huang}, {and} \bibinfo{person}{Hanghang Tong}} (Eds.).
  \bibinfo{publisher}{{ACM}}, \bibinfo{pages}{3828--3837}.
\newblock
\urldef\tempurl%
\url{https://doi.org/10.1145/3459637.3481953}
\showDOI{\tempurl}


\bibitem[Gu et~al\mbox{.}(2020)]%
        {DMT_JD}
\bibfield{author}{\bibinfo{person}{Yulong Gu}, \bibinfo{person}{Zhuoye Ding},
  \bibinfo{person}{Shuaiqiang Wang}, \bibinfo{person}{Lixin Zou},
  \bibinfo{person}{Yiding Liu}, {and} \bibinfo{person}{Dawei Yin}.}
  \bibinfo{year}{2020}\natexlab{}.
\newblock \showarticletitle{Deep Multifaceted Transformers for Multi-objective
  Ranking in Large-Scale E-commerce Recommender Systems}. In
  \bibinfo{booktitle}{\emph{{CIKM} '20: The 29th {ACM} International Conference
  on Information and Knowledge Management, Virtual Event, Ireland, October
  19-23, 2020}}, \bibfield{editor}{\bibinfo{person}{Mathieu d'Aquin},
  \bibinfo{person}{Stefan Dietze}, \bibinfo{person}{Claudia Hauff},
  \bibinfo{person}{Edward Curry}, {and} \bibinfo{person}{Philippe
  Cudr{\'{e}}{-}Mauroux}} (Eds.). \bibinfo{publisher}{{ACM}},
  \bibinfo{pages}{2493--2500}.
\newblock
\urldef\tempurl%
\url{https://doi.org/10.1145/3340531.3412697}
\showDOI{\tempurl}


\bibitem[He and McAuley(2016)]%
        {amazon_dataset}
\bibfield{author}{\bibinfo{person}{Ruining He} {and} \bibinfo{person}{Julian
  McAuley}.} \bibinfo{year}{2016}\natexlab{}.
\newblock \showarticletitle{Ups and Downs: Modeling the Visual Evolution of
  Fashion Trends with One-Class Collaborative Filtering}. In
  \bibinfo{booktitle}{\emph{Proceedings of the 25th International Conference on
  World Wide Web}} (Montr\'{e}al, Qu\'{e}bec, Canada)
  \emph{(\bibinfo{series}{WWW '16})}. \bibinfo{publisher}{International World
  Wide Web Conferences Steering Committee}, \bibinfo{address}{Republic and
  Canton of Geneva, CHE}, \bibinfo{pages}{507–517}.
\newblock
\showISBNx{9781450341431}
\urldef\tempurl%
\url{https://doi.org/10.1145/2872427.2883037}
\showDOI{\tempurl}


\bibitem[Jin et~al\mbox{.}(2020)]%
        {multi_behavior_SIGIR20}
\bibfield{author}{\bibinfo{person}{Bowen Jin}, \bibinfo{person}{Chen Gao},
  \bibinfo{person}{Xiangnan He}, \bibinfo{person}{Depeng Jin}, {and}
  \bibinfo{person}{Yong Li}.} \bibinfo{year}{2020}\natexlab{}.
\newblock \showarticletitle{Multi-Behavior Recommendation with Graph
  Convolutional Networks}. In \bibinfo{booktitle}{\emph{Proceedings of the 43rd
  International ACM SIGIR Conference on Research and Development in Information
  Retrieval}} (Virtual Event, China) \emph{(\bibinfo{series}{SIGIR '20})}.
  \bibinfo{publisher}{Association for Computing Machinery},
  \bibinfo{address}{New York, NY, USA}, \bibinfo{pages}{659–668}.
\newblock
\showISBNx{9781450380164}
\urldef\tempurl%
\url{https://doi.org/10.1145/3397271.3401072}
\showDOI{\tempurl}


\bibitem[Liu et~al\mbox{.}(2023)]%
        {liu2023JDsearch}
\bibfield{author}{\bibinfo{person}{Jiongnan Liu}, \bibinfo{person}{Zhicheng
  Dou}, \bibinfo{person}{Guoyu Tang}, {and} \bibinfo{person}{Sulong Xu}.}
  \bibinfo{year}{2023}\natexlab{}.
\newblock \showarticletitle{JDsearch: A Personalized Product Search Dataset
  with Real Queries and Full Interactions}. In
  \bibinfo{booktitle}{\emph{Proceedings of the {SIGIR} 2023}}.
  \bibinfo{publisher}{{ACM}}.
\newblock
\urldef\tempurl%
\url{https://doi.org/10.1145/3539618.3591900}
\showDOI{\tempurl}


\bibitem[Liu et~al\mbox{.}(2022)]%
        {Liu_www22}
\bibfield{author}{\bibinfo{person}{Jiongnan Liu}, \bibinfo{person}{Zhicheng
  Dou}, \bibinfo{person}{Qiannan Zhu}, {and} \bibinfo{person}{Ji-Rong Wen}.}
  \bibinfo{year}{2022}\natexlab{}.
\newblock \showarticletitle{A Category-Aware Multi-Interest Model for
  Personalized Product Search}. In \bibinfo{booktitle}{\emph{Proceedings of the
  ACM Web Conference 2022}} (Virtual Event, Lyon, France)
  \emph{(\bibinfo{series}{WWW '22})}. \bibinfo{publisher}{Association for
  Computing Machinery}, \bibinfo{address}{New York, NY, USA},
  \bibinfo{pages}{360–368}.
\newblock
\showISBNx{9781450390965}
\urldef\tempurl%
\url{https://doi.org/10.1145/3485447.3511964}
\showDOI{\tempurl}


\bibitem[Ma et~al\mbox{.}(2018)]%
        {MMOE}
\bibfield{author}{\bibinfo{person}{Jiaqi Ma}, \bibinfo{person}{Zhe Zhao},
  \bibinfo{person}{Xinyang Yi}, \bibinfo{person}{Jilin Chen},
  \bibinfo{person}{Lichan Hong}, {and} \bibinfo{person}{Ed~H. Chi}.}
  \bibinfo{year}{2018}\natexlab{}.
\newblock \showarticletitle{Modeling Task Relationships in Multi-task Learning
  with Multi-gate Mixture-of-Experts}. In \bibinfo{booktitle}{\emph{Proceedings
  of the 24th {ACM} {SIGKDD} International Conference on Knowledge Discovery
  {\&} Data Mining, {KDD} 2018, London, UK, August 19-23, 2018}},
  \bibfield{editor}{\bibinfo{person}{Yike Guo} {and} \bibinfo{person}{Faisal
  Farooq}} (Eds.). \bibinfo{publisher}{{ACM}}, \bibinfo{pages}{1930--1939}.
\newblock
\urldef\tempurl%
\url{https://doi.org/10.1145/3219819.3220007}
\showDOI{\tempurl}


\bibitem[Si et~al\mbox{.}(2022)]%
        {IV4REC}
\bibfield{author}{\bibinfo{person}{Zihua Si}, \bibinfo{person}{Xueran Han},
  \bibinfo{person}{Xiao Zhang}, \bibinfo{person}{Jun Xu}, \bibinfo{person}{Yue
  Yin}, \bibinfo{person}{Yang Song}, {and} \bibinfo{person}{Ji-Rong Wen}.}
  \bibinfo{year}{2022}\natexlab{}.
\newblock \showarticletitle{A Model-Agnostic Causal Learning Framework for
  Recommendation Using Search Data}. In \bibinfo{booktitle}{\emph{Proceedings
  of the ACM Web Conference 2022}} (Virtual Event, Lyon, France)
  \emph{(\bibinfo{series}{WWW '22})}. \bibinfo{publisher}{Association for
  Computing Machinery}, \bibinfo{address}{New York, NY, USA},
  \bibinfo{pages}{224–233}.
\newblock
\showISBNx{9781450390965}
\urldef\tempurl%
\url{https://doi.org/10.1145/3485447.3511951}
\showDOI{\tempurl}


\bibitem[Si et~al\mbox{.}(2023a)]%
        {IV4Rec+}
\bibfield{author}{\bibinfo{person}{Zihua Si}, \bibinfo{person}{Zhongxiang Sun},
  \bibinfo{person}{Xiao Zhang}, \bibinfo{person}{Jun Xu}, \bibinfo{person}{Yang
  Song}, \bibinfo{person}{Xiaoxue Zang}, {and} \bibinfo{person}{Ji-Rong Wen}.}
  \bibinfo{year}{2023}\natexlab{a}.
\newblock \showarticletitle{Enhancing Recommendation with Search Data in a
  Causal Learning Manner}.
\newblock \bibinfo{journal}{\emph{ACM Transactions on Information Systems}}
  (\bibinfo{date}{Feb} \bibinfo{year}{2023}).
\newblock
\showISSN{1046-8188}
\urldef\tempurl%
\url{https://doi.org/10.1145/3582425}
\showDOI{\tempurl}


\bibitem[Si et~al\mbox{.}(2023b)]%
        {SESRec}
\bibfield{author}{\bibinfo{person}{Zihua Si}, \bibinfo{person}{Zhongxiang Sun},
  \bibinfo{person}{Xiao Zhang}, \bibinfo{person}{Jun Xu},
  \bibinfo{person}{Xiaoxue Zang}, \bibinfo{person}{Yang Song},
  \bibinfo{person}{Kun Gai}, {and} \bibinfo{person}{Ji-Rong Wen}.}
  \bibinfo{year}{2023}\natexlab{b}.
\newblock \showarticletitle{When Search Meets Recommendation: Learning
  Disentangled Search Representation for Recommendation}. In
  \bibinfo{booktitle}{\emph{Proceedings of the 46th International ACM SIGIR
  Conference on Research and Development in Information Retrieval}} (Taipei,
  Taiwan) \emph{(\bibinfo{series}{SIGIR '23})}. \bibinfo{publisher}{Association
  for Computing Machinery}, \bibinfo{address}{New York, NY, USA},
  \bibinfo{pages}{1313–1323}.
\newblock
\showISBNx{9781450394086}
\urldef\tempurl%
\url{https://doi.org/10.1145/3539618.3591786}
\showDOI{\tempurl}


\bibitem[Tang et~al\mbox{.}(2020)]%
        {PLE}
\bibfield{author}{\bibinfo{person}{Hongyan Tang}, \bibinfo{person}{Junning
  Liu}, \bibinfo{person}{Ming Zhao}, {and} \bibinfo{person}{Xudong Gong}.}
  \bibinfo{year}{2020}\natexlab{}.
\newblock \showarticletitle{Progressive Layered Extraction {(PLE):} {A} Novel
  Multi-Task Learning {(MTL)} Model for Personalized Recommendations}. In
  \bibinfo{booktitle}{\emph{RecSys 2020: Fourteenth {ACM} Conference on
  Recommender Systems, Virtual Event, Brazil, September 22-26, 2020}},
  \bibfield{editor}{\bibinfo{person}{Rodrygo L.~T. Santos},
  \bibinfo{person}{Leandro~Balby Marinho}, \bibinfo{person}{Elizabeth~M. Daly},
  \bibinfo{person}{Li~Chen}, \bibinfo{person}{Kim Falk}, \bibinfo{person}{Noam
  Koenigstein}, {and} \bibinfo{person}{Edleno~Silva de~Moura}} (Eds.).
  \bibinfo{publisher}{{ACM}}, \bibinfo{pages}{269--278}.
\newblock
\urldef\tempurl%
\url{https://doi.org/10.1145/3383313.3412236}
\showDOI{\tempurl}


\bibitem[Wu et~al\mbox{.}(2019)]%
        {NRHUB}
\bibfield{author}{\bibinfo{person}{Chuhan Wu}, \bibinfo{person}{Fangzhao Wu},
  \bibinfo{person}{Mingxiao An}, \bibinfo{person}{Tao Qi},
  \bibinfo{person}{Jianqiang Huang}, \bibinfo{person}{Yongfeng Huang}, {and}
  \bibinfo{person}{Xing Xie}.} \bibinfo{year}{2019}\natexlab{}.
\newblock \showarticletitle{Neural News Recommendation with Heterogeneous User
  Behavior}. In \bibinfo{booktitle}{\emph{Proceedings of the 2019 Conference on
  Empirical Methods in Natural Language Processing and the 9th International
  Joint Conference on Natural Language Processing (EMNLP-IJCNLP)}}.
  \bibinfo{publisher}{Association for Computational Linguistics},
  \bibinfo{address}{Hong Kong, China}, \bibinfo{pages}{4874--4883}.
\newblock
\urldef\tempurl%
\url{https://doi.org/10.18653/v1/D19-1493}
\showDOI{\tempurl}


\bibitem[Wu et~al\mbox{.}(2022)]%
        {multi_behavior_www22}
\bibfield{author}{\bibinfo{person}{Chuhan Wu}, \bibinfo{person}{Fangzhao Wu},
  \bibinfo{person}{Tao Qi}, \bibinfo{person}{Qi Liu}, \bibinfo{person}{Xuan
  Tian}, \bibinfo{person}{Jie Li}, \bibinfo{person}{Wei He},
  \bibinfo{person}{Yongfeng Huang}, {and} \bibinfo{person}{Xing Xie}.}
  \bibinfo{year}{2022}\natexlab{}.
\newblock \showarticletitle{FeedRec: News Feed Recommendation with Various User
  Feedbacks}. In \bibinfo{booktitle}{\emph{Proceedings of the ACM Web
  Conference 2022}} (Virtual Event, Lyon, France) \emph{(\bibinfo{series}{WWW
  '22})}. \bibinfo{publisher}{Association for Computing Machinery},
  \bibinfo{address}{New York, NY, USA}, \bibinfo{pages}{2088–2097}.
\newblock
\showISBNx{9781450390965}
\urldef\tempurl%
\url{https://doi.org/10.1145/3485447.3512082}
\showDOI{\tempurl}


\bibitem[Wu et~al\mbox{.}(2020)]%
        {Google_trasnfer_from_src_to_reco}
\bibfield{author}{\bibinfo{person}{Tao Wu}, \bibinfo{person}{Ellie Ka~In Chio},
  \bibinfo{person}{Heng{-}Tze Cheng}, \bibinfo{person}{Yu Du},
  \bibinfo{person}{Steffen Rendle}, \bibinfo{person}{Dima Kuzmin},
  \bibinfo{person}{Ritesh Agarwal}, \bibinfo{person}{Li Zhang},
  \bibinfo{person}{John~R. Anderson}, \bibinfo{person}{Sarvjeet Singh},
  \bibinfo{person}{Tushar Chandra}, \bibinfo{person}{Ed~H. Chi},
  \bibinfo{person}{Wen Li}, \bibinfo{person}{Ankit Kumar},
  \bibinfo{person}{Xiang Ma}, \bibinfo{person}{Alex Soares},
  \bibinfo{person}{Nitin Jindal}, {and} \bibinfo{person}{Pei Cao}.}
  \bibinfo{year}{2020}\natexlab{}.
\newblock \showarticletitle{Zero-Shot Heterogeneous Transfer Learning from
  Recommender Systems to Cold-Start Search Retrieval}. In
  \bibinfo{booktitle}{\emph{{CIKM} '20: The 29th {ACM} International Conference
  on Information and Knowledge Management, Virtual Event, Ireland, October
  19-23, 2020}}, \bibfield{editor}{\bibinfo{person}{Mathieu d'Aquin},
  \bibinfo{person}{Stefan Dietze}, \bibinfo{person}{Claudia Hauff},
  \bibinfo{person}{Edward Curry}, {and} \bibinfo{person}{Philippe
  Cudr{\'{e}}{-}Mauroux}} (Eds.). \bibinfo{publisher}{{ACM}},
  \bibinfo{pages}{2821--2828}.
\newblock
\urldef\tempurl%
\url{https://doi.org/10.1145/3340531.3412752}
\showDOI{\tempurl}


\bibitem[Xiong et~al\mbox{.}(2019)]%
        {Xiong2019OpenDW}
\bibfield{author}{\bibinfo{person}{Lee Xiong}, \bibinfo{person}{Chuan Hu},
  \bibinfo{person}{Chenyan Xiong}, \bibinfo{person}{Daniel~Fernando Campos},
  {and} \bibinfo{person}{Arnold Overwijk}.} \bibinfo{year}{2019}\natexlab{}.
\newblock \showarticletitle{Open Domain Web Keyphrase Extraction Beyond
  Language Modeling}. In \bibinfo{booktitle}{\emph{Conference on Empirical
  Methods in Natural Language Processing}}.
\newblock


\bibitem[Xu et~al\mbox{.}(2018)]%
        {xu2018deep}
\bibfield{author}{\bibinfo{person}{Jun Xu}, \bibinfo{person}{Xiangnan He},
  {and} \bibinfo{person}{Hang Li}.} \bibinfo{year}{2018}\natexlab{}.
\newblock \showarticletitle{Deep learning for matching in search and
  recommendation}. In \bibinfo{booktitle}{\emph{The 41st International ACM
  SIGIR Conference on Research \& Development in Information Retrieval}}.
  \bibinfo{pages}{1365--1368}.
\newblock


\bibitem[Yang et~al\mbox{.}(2021)]%
        {intent_reco_www21}
\bibfield{author}{\bibinfo{person}{Yatao Yang}, \bibinfo{person}{Biyu Ma},
  \bibinfo{person}{Jun Tan}, \bibinfo{person}{Hongbo Deng},
  \bibinfo{person}{Haikuan Huang}, {and} \bibinfo{person}{Zibin Zheng}.}
  \bibinfo{year}{2021}\natexlab{}.
\newblock \showarticletitle{FINN: Feedback Interactive Neural Network for
  Intent Recommendation}. In \bibinfo{booktitle}{\emph{Proceedings of the Web
  Conference 2021}} (Ljubljana, Slovenia) \emph{(\bibinfo{series}{WWW '21})}.
  \bibinfo{publisher}{Association for Computing Machinery},
  \bibinfo{address}{New York, NY, USA}, \bibinfo{pages}{1949–1958}.
\newblock
\showISBNx{9781450383127}
\urldef\tempurl%
\url{https://doi.org/10.1145/3442381.3450105}
\showDOI{\tempurl}


\bibitem[Yao et~al\mbox{.}(2021)]%
        {USER}
\bibfield{author}{\bibinfo{person}{Jing Yao}, \bibinfo{person}{Zhicheng Dou},
  \bibinfo{person}{Ruobing Xie}, \bibinfo{person}{Yanxiong Lu},
  \bibinfo{person}{Zhiping Wang}, {and} \bibinfo{person}{Ji-Rong Wen}.}
  \bibinfo{year}{2021}\natexlab{}.
\newblock \showarticletitle{USER: A Unified Information Search and
  Recommendation Model Based on Integrated Behavior Sequence}. In
  \bibinfo{booktitle}{\emph{Proceedings of the 30th ACM International
  Conference on Information ]\&amp; Knowledge Management}} (Virtual Event,
  Queensland, Australia) \emph{(\bibinfo{series}{CIKM '21})}.
  \bibinfo{publisher}{Association for Computing Machinery},
  \bibinfo{address}{New York, NY, USA}, \bibinfo{pages}{2373–2382}.
\newblock
\showISBNx{9781450384469}
\urldef\tempurl%
\url{https://doi.org/10.1145/3459637.3482489}
\showDOI{\tempurl}


\bibitem[Yuan et~al\mbox{.}(2022)]%
        {Multi_behavior_SIGIR22}
\bibfield{author}{\bibinfo{person}{Enming Yuan}, \bibinfo{person}{Wei Guo},
  \bibinfo{person}{Zhicheng He}, \bibinfo{person}{Huifeng Guo},
  \bibinfo{person}{Chengkai Liu}, {and} \bibinfo{person}{Ruiming Tang}.}
  \bibinfo{year}{2022}\natexlab{}.
\newblock \showarticletitle{Multi-Behavior Sequential Transformer Recommender}
  \emph{(\bibinfo{series}{SIGIR '22})}. \bibinfo{publisher}{Association for
  Computing Machinery}, \bibinfo{address}{New York, NY, USA},
  \bibinfo{pages}{1642–1652}.
\newblock
\showISBNx{9781450387323}
\urldef\tempurl%
\url{https://doi.org/10.1145/3477495.3532023}
\showDOI{\tempurl}


\bibitem[Zamani and Croft(2018)]%
        {JSR}
\bibfield{author}{\bibinfo{person}{Hamed Zamani} {and}
  \bibinfo{person}{W.~Bruce Croft}.} \bibinfo{year}{2018}\natexlab{}.
\newblock \showarticletitle{Joint Modeling and Optimization of Search and
  Recommendation}. In \bibinfo{booktitle}{\emph{Proceedings of the First
  Biennial Conference on Design of Experimental Search {\&} Information
  Retrieval Systems, Bertinoro, Italy, August 28-31, 2018}}
  \emph{(\bibinfo{series}{{CEUR} Workshop Proceedings},
  Vol.~\bibinfo{volume}{2167})}. \bibinfo{publisher}{CEUR-WS.org},
  \bibinfo{pages}{36--41}.
\newblock


\bibitem[Zamani and Croft(2020)]%
        {JSR2}
\bibfield{author}{\bibinfo{person}{Hamed Zamani} {and}
  \bibinfo{person}{W.~Bruce Croft}.} \bibinfo{year}{2020}\natexlab{}.
\newblock \showarticletitle{Learning a Joint Search and Recommendation Model
  from User-Item Interactions}. In \bibinfo{booktitle}{\emph{Proceedings of the
  13th International Conference on Web Search and Data Mining}} (Houston, TX,
  USA) \emph{(\bibinfo{series}{WSDM '20})}. \bibinfo{publisher}{Association for
  Computing Machinery}, \bibinfo{address}{New York, NY, USA},
  \bibinfo{pages}{717–725}.
\newblock
\showISBNx{9781450368223}
\urldef\tempurl%
\url{https://doi.org/10.1145/3336191.3371818}
\showDOI{\tempurl}


\bibitem[Zhang et~al\mbox{.}(2021)]%
        {Zhang2021Counterfactual}
\bibfield{author}{\bibinfo{person}{Xiao Zhang}, \bibinfo{person}{Haonan Jia},
  \bibinfo{person}{Hanjing Su}, \bibinfo{person}{Wenhan Wang},
  \bibinfo{person}{Jun Xu}, {and} \bibinfo{person}{Ji{-}Rong Wen}.}
  \bibinfo{year}{2021}\natexlab{}.
\newblock \showarticletitle{Counterfactual reward modification for streaming
  recommendation with delayed feedback}. In
  \bibinfo{booktitle}{\emph{Proceedings of the 44th International {ACM} {SIGIR}
  Conference on Research and Development in Information Retrieval}}.
  \bibinfo{pages}{41--50}.
\newblock


\bibitem[Zhao et~al\mbox{.}(2022)]%
        {SRJgraph}
\bibfield{author}{\bibinfo{person}{Kai Zhao}, \bibinfo{person}{Yukun Zheng},
  \bibinfo{person}{Tao Zhuang}, \bibinfo{person}{Xiang Li}, {and}
  \bibinfo{person}{Xiaoyi Zeng}.} \bibinfo{year}{2022}\natexlab{}.
\newblock \showarticletitle{Joint Learning of E-Commerce Search and
  Recommendation with a Unified Graph Neural Network}. In
  \bibinfo{booktitle}{\emph{Proceedings of the Fifteenth ACM International
  Conference on Web Search and Data Mining}} (Virtual Event, AZ, USA)
  \emph{(\bibinfo{series}{WSDM '22})}. \bibinfo{publisher}{Association for
  Computing Machinery}, \bibinfo{address}{New York, NY, USA},
  \bibinfo{pages}{1461–1469}.
\newblock
\showISBNx{9781450391320}
\urldef\tempurl%
\url{https://doi.org/10.1145/3488560.3498414}
\showDOI{\tempurl}


\bibitem[Zou et~al\mbox{.}(2022)]%
        {muliti_scenario_SIGIR22}
\bibfield{author}{\bibinfo{person}{Xinyu Zou}, \bibinfo{person}{Zhi Hu},
  \bibinfo{person}{Yiming Zhao}, \bibinfo{person}{Xuchu Ding},
  \bibinfo{person}{Zhongyi Liu}, \bibinfo{person}{Chenliang Li}, {and}
  \bibinfo{person}{Aixin Sun}.} \bibinfo{year}{2022}\natexlab{}.
\newblock \showarticletitle{Automatic Expert Selection for Multi-Scenario and
  Multi-Task Search}. In \bibinfo{booktitle}{\emph{{SIGIR} '22: The 45th
  International {ACM} {SIGIR} Conference on Research and Development in
  Information Retrieval, Madrid, Spain, July 11 - 15, 2022}},
  \bibfield{editor}{\bibinfo{person}{Enrique Amig{\'{o}}},
  \bibinfo{person}{Pablo Castells}, \bibinfo{person}{Julio Gonzalo},
  \bibinfo{person}{Ben Carterette}, \bibinfo{person}{J.~Shane Culpepper}, {and}
  \bibinfo{person}{Gabriella Kazai}} (Eds.). \bibinfo{publisher}{{ACM}},
  \bibinfo{pages}{1535--1544}.
\newblock
\urldef\tempurl%
\url{https://doi.org/10.1145/3477495.3531942}
\showDOI{\tempurl}


\end{thebibliography}

\appendix

\end{document}